# Discrete nature of inhomogeneity: The initial stages and local configurations of TiOPc during bilayer growth on Ag(111).

Laura Fernandez[a], Sebastian Thussing[a], Alexander Mänz[a], Jörg Sundermeyer[b], Gregor Witte[a] and Peter Jakob [a]*

[a] Fachbereich Physik, Philipps-Universität Marburg, Renthof 5, 35032 Marburg, Germany.
[b] Fachbereich Chemie, Philipps-Universität Marburg, Hans-Meerweinstr. 4, 35032 Marburg, Germany.

**Abstract**

The operation of organic optoelectronic devices relies notably on the bulk properties of compound molecular species, but even more so on the influence of interfaces thereof. The identification and characterization of elemental processes in these critical sections of a device thereby requires well-defined interfaces with low defect density. In this context titanyl phthalocyanine (TiOPc) arises as an excellent candidate that reveals the formation of a stable bilayer structure with a characteristic "up-down" molecular arrangement that optimizes the dipole-dipole interaction within the bilayer. In our experimental study, long-range ordered TiOPc bilayers have been grown on Ag(111) surfaces and analyzed using infrared absorption spectroscopy and scanning tunneling microscopy. By monitoring the prominent Ti=O stretching mode in IRAS and identifying local configurations in STM, a microscopic model for the growth of TiOPc bilayers on Ag(111) is suggested. We demonstrate that defect structures within these bilayers lead to characteristic vibrational signatures which react sensitively to the local environment of the molecules. Thermal desorption spectroscopy reveals a high thermal stability of the TiOPc bilayer up to 500 K, which is attributed to hydrogen bonds between oxygen of the titanyl unit and the hydrogen rim of phthalocyanines in the second layer, in addition to contributions arising from the oppositely oriented axial dipole moments and the ubiquitous van der Waals interactions.

*E-Mail: peter.jakob@physik.uni-marburg.de

**Introduction**

The quality and versatility of organic optoelectronic devices is intimately connected with the design and availability of new functional materials with photoactive properties as well as their controlled microstructural assembly. Hereby the interfaces between organic layers and to the substrate represent key elements of associated hetero-organic layers. Of particular interest in this regard are interfaces between two molecular species where complex exciton dissociation processes take place,[1] or between organic films and (metallic) electrodes as they facilitate the electric contacting.[2,3] In order to identify elementary steps in their operation and to develop microscopic models, it is essential to grow well-ordered and homogeneous molecular films with tailored electronic properties and preferably smooth interfaces between adjacent organic layers. Therefore, defect control is a key objective and requires sensitive methods capable of characterizing defects in crystalline organic thin films. In contrast to covalently or ionic crystals (such as e.g. alkali halides or oxides) where vacancies exhibit characteristic optical signatures (known as F-centers),[4] such features are hardly found in van der Waals bound crystalline organic semiconductors. Though scanning probe microscopy has provided very detailed insight into defects in molecular films, this information is limited to the outermost layer. By contrast, high resolution transmission electron microscopy also allows visualization of defects in buried interfaces; however, the high radiation sensitivity of molecular materials as well as special requirements on acceptable sample thickness render this method a rather invasive approach whose sensitivity also largely suffers from the low Z-contrast of hydrocarbons.[5]

Here, we demonstrate that vibrational signatures of molecules are particularly sensitive to the local environment and thus can be used to identify and characterize differently coordinated vacancies. In this regard the nonplanar titanyl phthalocyanine (TiOPc), with its intrinsic electric dipole moment associated with an axial Ti=O group, appears as an interesting candidate since the Ti=O stretching mode reveal a large oscillator strength. Previous experimental works have demonstrated the ability of TiOPc molecules to form ordered multilayer films, where the high dipole moment of the titanyl group Ti=O is thought to play an important role in the formation of homogeneous layers and bilayers.[6-8] A particularity of TiOPc, namely its shallow square pyramidal structure, leads to the existence of numerous polymorphs with different stacking modes and molecular alignments in bulk materials as well as grown solid films each displaying characteristic conductivities, hole mobilities and optical properties.[9-11]

In the present work we have performed at first an in-depth investigation of the growth mechanism and structure of TiOPc bilayers on a Ag(111) surface. By means of spot profile analysis low energy electron diffraction (SPA-LEED), scanning tunneling microscopy (STM), and infrared absorption spectroscopy (IRAS) we are able to develop a microscopic model regarding the arrangement and orientation of TiOPc molecules in mono- and bilayer films on Ag(111). Particularly appealing is the stability found for the TiOPc bilayer, as demonstrated by the high desorption temperature of around 550 K measured by thermal desorption spectroscopy (TDS). By correlating IRAS and STM data we are able to establish a correlation between the characteristic vibrational signatures of the titanyl stretch mode with structural patterns, i.e. local inhomogeneities in the molecular film, occurring upon completion of the bilayer.

**Experimental**

The measurements reported here were performed in two different ultrahigh vacuum (UHV) systems. Spot-profile analysis low energy electron diffraction (SPA-LEED, Omicron, thermal desorption spectroscopy and Fourier transform infrared absorption spectroscopy (FT-IRAS, Bruker IFS 66v/s) with evacuated optics (p < 10 mbar) were performed in one system described in detail elsewhere.[12] IRAS measurements were performed at a sample temperature of 80 K and, typically, 1000-2000 scans have been co-added at 2 cm$^{-1}$ spectral resolution. The STM measurements were obtained in another system using a variable temperature scanning tunneling microscope (Omicron VT-STM) operated at 110 K in the constant current mode and using etched tungsten tips. For the present study, different silver substrates have been used: a Ag(111) single crystal (Mateck, purity 5N), as well as Ag(111) films (150 nm) that were epitaxially grown under high vacuum conditions onto freshly cleaved and carefully degassed mica substrates. Prior to each organic film deposition, the silver surfaces were cleaned in situ by repeated cycles of Ar$^+$ sputtering (E ≈ 800 eV) and annealing (≈750 K) until a sharp (1x1)-LEED pattern with a low background signal was observed and no traces of contaminations were found by STM or IRAS. The temperature of the substrates was measured using a K-type thermocouple attached directly to the sample, hence providing a precise temperature reading. TiOPc purified by extensive degassing at 550 K and analyzed by quadrupole mass spectrometry) thin films were prepared under UHV conditions by sublimation from a Knudsen cell. To facilitate formation of highly ordered films, deposition was performed at non-cryogenic substrate temperatures (300 - 450 K) and using a low deposition rate of 0.05 ML/min. The deposition rate was estimated from IRAS measurements which reveal a clear signature of monolayer (ML) completion, that is, saturation of monolayer bands and appearance of distinct bilayer vibrational modes. For the STM measurements TiOPc films were grown at typical deposition rates of 0.5 Å/min (0.2 ML/min), as determined by a quartz crystal microbalance and STM. The desorption traces of the TiOPc bilayers and multilayers were investigated by TDS. Thereby, several masses of our quadrupole mass spectrometer (Pfeiffer, QMG 700, mass range 0–1024 u) were collected 'in parallel' while applying a linear heating rate of 1 K/s.

## Results and Discussion

### A  Structure of the TiOPc mono- and bilayer

The TiOPc coverage was systematically increased from 1ML till completion of the second layer. The full monolayer of TiOPc on Ag(111) is characterized by a point-on-line (POL) phase which is readily identified by SPA-LEED and STM measurements (cf. Fig. 1).[18] The STM data reveal a uniform azimuthal orientation of the TiOPc backbone plane with respect to the substrate azimuth direction which avoids an overlap of neighbouring molecules and thus allows to minimize repulsive interactions among the closely spaced molecules.[6,13-18] A detailed analysis of the LEED pattern reveals that the POL-phase of the TiOPc monolayer in Figure 1a can actually be classified as a high order commensurate (HOC) phase and described by the matrix $\begin{pmatrix} 4.875 & -0.125 \\ 2.625 & 5.625 \end{pmatrix}$ where a 2D registry between molecules and the Ag(111) surface is only partially lost. The corresponding unit cell, indicated in red (Figure 1a) has a size of 199 Å$^2$ and contains only a single molecule.

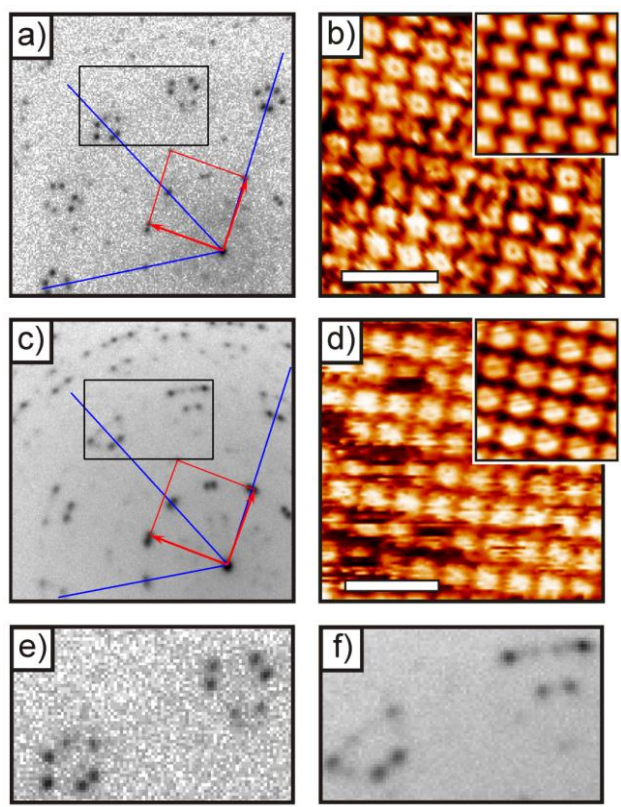

**Figure 1** SPA-LEED images and corresponding STM micrographs of the TiOPc mono- and bilayer on Ag(111). (a,c) refer to SPA-LEED measurements of the TiOPc/Ag(111) saturated monolayer, and to the TiOPc bilayer structure on Ag(111), respectively; the TiOPc surface unit cell and <1-10> directions of Ag(111) are indicated in red and blue, respectively. (b,d) The STM micrographs reveal that the surface unit cell contains only one molecule for both, the mono- and the bilayer structure; inset micrographs represent correlated averages. (e, f) are enlargements of the LEED patterns in (a) and (c), respectively, as indicated by the black box. SPA-LEED measurements were obtained at 80 K ($E_{kin}$ = 50 eV). STM micrographs were taken at T = 110 K ($U_{Bias}$ = 0.43 V, $I_t$ = 10 pA). The scale bars in (b) and (d) amount to 5 nm.

A further increase of the coverage beyond this value, which is set to 1 ML, leads to the growth of TiOPc molecules in a second layer. Since the second-layer molecules are no longer in direct contact with the metal surface they experience a weaker substrate interaction as compared to the first-layer molecules. This is confirmed by our TDS data which yield virtually complete desorption of second layer TiOPc, while monolayer species adsorb intact and dissociate at elevated T > 600 K. A central question regarding the thin film growth of TiOPc therefore concerns the molecular arrangement of the second layer species and whether it differs from those found in the first layer. It is expected that the intermolecular interaction between adjacent layers and neighbouring molecules will play an essential role here.

A characteristic LEED pattern obtained after completion of the second TiOPc layer is displayed in Fig. 1c. Though the close-to-rectangular unit cell of the TiOPc/Ag(111) monolayer is maintained, it is clearly different from the POL-phase of the first layer which reflects some molecular rearrangements upon addition of a second layer. In matrix notation this bilayer structure can be described by $\begin{pmatrix} 4.88 & 0.19 \\ 2.56 & 5.50 \end{pmatrix}$. All matrix descriptions of the observed overlayer structures have been verified by a proper modelling and reproduction of the respective LEED patterns using the LEEDpat simulation program[19]. The unit cell area of the TiOPc bilayer amounts to 189 Å$^2$ which is notably smaller than the unit cell size of the POL-phase. Interestingly, the LEED pattern of the completed bilayer reveals no residual reflexes from the first layer underneath. This indicates an additional compression of the TiOPc monolayer after deposition of the second molecular layer.

Accompanied STM data (cf. Fig. 1c) again reveal a uniform azimuthal alignment with a nearly rectangular unit cell which suggests a primitive unit cell. We note that imaging of the not-yet-completed TiOPc bilayer is impeded by weakly bound, diffusing admolecules.

To derive additional information on the orientation of the titanyl group complementary IRAS measurements have been performed. Fig. 2 shows an overview of the vibrational spectra associated with the full monolayer (POL-phase) and bilayer of TiOPc on Ag(111). The spectral region with the most prominent bands, i.e. 700 - 1000 cm$^{-1}$, primarily covers modes with out-of-plane character (this description uses the π-conjugated backbone of TiOPc as the reference plane). Among them, the two C-H bending modes (714.5 and 763.8 cm$^{-1}$), located at the periphery of the Pc molecular frame, and the axial Ti=O stretching mode at 993.3 cm$^{-1}$ are readily identified.[20]

In the bilayer spectrum the two C-H bending modes of first-layer TiOPc have lost some intensity and they have shifted to slightly different line positions (influenced by the extra TiOPc layer on top). Of course, the bilayer spectrum reveals extra features due to second-layer TiOPc, e.g. the two intense bands at 739 and 784.5 cm$^{-1}$ that can be ascribed to the C-H bending modes mentioned above; their vibrational frequencies are now much closer to the bulk values (≈735 and ≈785 cm$^{-1}$, cf. Fig. 2a and Fig. 6) due to the vanishing influence of the Ag substrate. Interestingly, the Ti=O stretching mode (located at 993.3 cm$^{-1}$ for the POL monolayer phase), exhibits a noticeable frequency shift to 956.9 cm$^{-1}$ upon completion of the bilayer. This softening of the Ti=O stretch mode seems directly related to the characteristic arrangement of the TiOPc molecules within the second layer and reflects specific interactions between first- and second-layer molecules. This aspect will be discussed in more detail in section B.

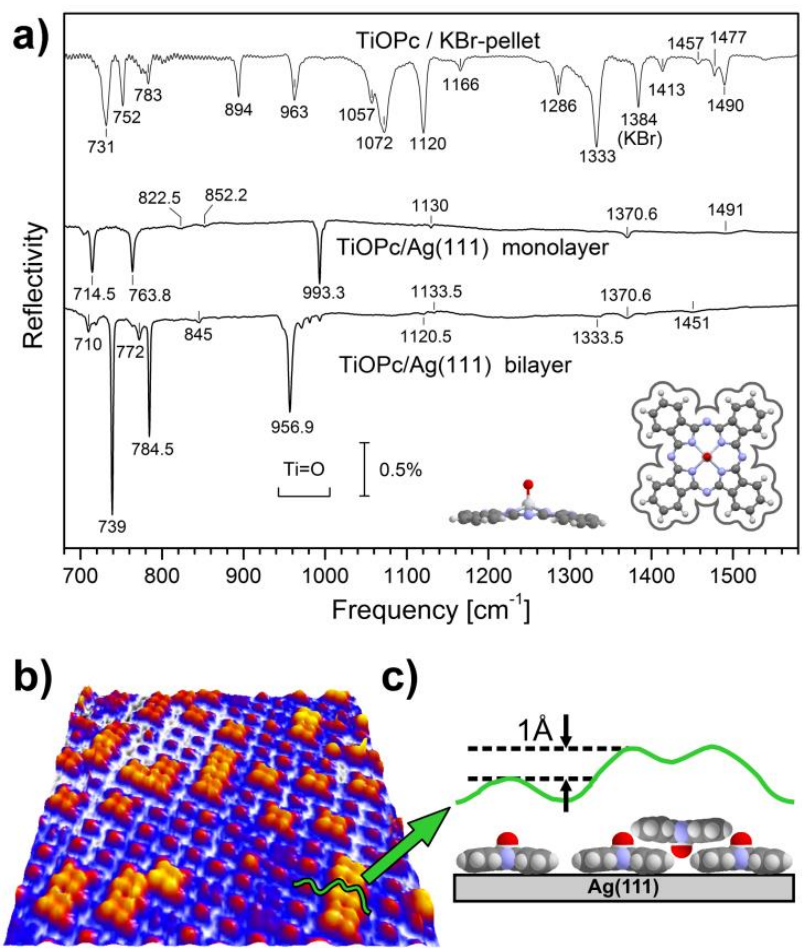

**Figure 2** (a) Comparison of IR absorption spectra of TiOPc adsorbed on Ag(111). The middle spectrum refers to one full monolayer (POL-phase); the bottom spectrum has been obtained from a TiOPc bilayer. For comparison the TiOPc/KBr-pellet spectrum is shown (top curve). The displayed spectral range covers out-of-plane vibrational modes (C-H bending modes and Ti=O stretching mode) at <1000cm$^{-1}$, as well as weak in-plane modes at higher frequencies.[20] Discrimination of mono- and bilayer TiOPc species is straightforward from the respective distinct vibrational features. The spectra have been obtained at 80 K using a spectral resolution of 2 cm$^{-1}$. (b) STM micrograph ($U_{Bias}$ = 0.57 V, $I_t$ = 30 pA) of 1.2 ML TiOPc on Ag(111) showing the presence of bilayer molecules with a distinct four-lobed shape while first-layer molecules exhibit one central protrusion that is associated with the titanyl unit. (c) The corresponding line scan reveals a height difference between first layer protrusions and the aromatic frame of second layer molecules of about 1Å which is in line with the depicted interdigitated stacking.

In the following, we will focus on the 'in-plane' vibrational modes that are detected above 1000 cm$^{-1}$ in Fig. 2a, and which are essential to extract information regarding the molecular orientation. The term 'in-plane' thereby assumes that the phthalocyanine backbone represents a virtually flat entity, even though the TiOPc molecule shows a slight pyramidal shape as shown in the inset of Fig. 2a. This non-planarity renders A$_1$-type of modes dipole active, even for parallel adsorption geometries, with the conjugated backbone oriented parallel to the surface, as well as for configurations with negligible electronic coupling to the metal substrate. For molecules grown on metallic surfaces, like in the present case, only those vibrations with dynamic dipole moments perpendicular to the metal surface (out-of-plane and totally symmetric in-plane vibrational modes for parallel oriented molecules) are detected by IRAS due to the surface selection rule.[21]

According to Fig. 2, weak in-plane vibrational modes, some with an asymmetric line shape, are observed at about 1130, 1370.6 and 1491 cm$^{-1}$ for the saturated TiOPc/Ag(111) monolayer (POL-phase). They are ascribed to in-plane deformation modes of the pyrrole and the benzene rings with an A$_1$ - type of symmetry. These modes may acquire substantial intensity due to interfacial dynamical charge transfer (IDCT) induced by vibrational motion.[22-24] Asymmetric line shapes then are clear indications for non-adiabatic processes governing these excitations and the associated coupling between vibrational and electronic degrees of freedom. For the 1370.6 and 1491 cm$^{-1}$ bands this asymmetry is clearly discernible, even though it is not quite as prominent as compared to CuPc/Ag(111) and their intensities are much lower.[12]

Apart from these weak vibrational bands, no further prominent in-plane modes are detected. According to vibrational spectra of TiOPc/KBr-pellet samples (top curve in Fig. 2a or thick evaporated films,[9] very prominent bands with E-type of symmetry (irreducible representation of the C$_{4v}$ symmetry group) do exist in the spectral region 1000 - 1600 cm$^{-1}$. Their absence in the spectra of the grown films in Fig. 2a, in conjunction with the observed dominance of out-of-plane modes, unambiguously demonstrates that TiOPc has its molecular plane oriented parallel to the Ag(111) surface in the first ML.

In the case of the bilayer spectrum additional (again rather weak) in-plane modes are found at 1120.5, 1333.5, and 1451 cm$^{-1}$. Interestingly, all of them are located at slightly lower frequencies as compared to similar bands of the monolayer species. This indicates that they correspond to the same type of vibrational mode of second-layer TiOPc. For these species the frequency shifts experienced by the first monolayer due to direct contact to Ag(111) becomes attenuated and the observed line positions are then much closer to the vibrational energies of the free molecule or in the crystalline bulk (pellet sample). The weak intensity of these in-plane modes is again taken as a clear indication that the molecular frame of TiOPc molecules grown in the second layer is not (or only very slightly) inclined with respect to the Ag(111) metal surface. Note that the weak but clearly discernible intensities of vibrational modes in the 1000 - 1600 cm$^{-1}$ spectral region most likely are due to the slight pyramidal shape of the π- conjugated backbone of TiOPc.

The dissimilar molecular arrangement with the titanyl group oriented up and down in the TiOPc mono- and bilayers could actually be observed directly in some STM images (Figure 2b). Thereby a distinct submolecular contrast is discernible. While the molecules appear as quadratic objects for a wide range of tunnelling voltages (cf. Fig. 1), the oxygen atoms in the center of first-layer TiOPc molecules in Fig. 2b are imaged as distinct protrusions. We attribute this particular appearance to a very sharp tip or a molecule attached to the tip. By contrast, the simultaneously imaged TiOPc molecules of the second layer appear as bright four-lobed units, which are associated with the conjugated backbone. Interestingly, these bilayer molecules are arranged in groups of one, two, three, or more molecules.

From X-ray diffraction data of TiOPc bulk samples it is known that molecules arranges in a face-to-face manner forming bilayer stacks of oppositely oriented titanyl units. The most prominent polymorphs, denoted as phase I and phase II,[9] exhibit configurations where the second-layer TiOPc is located in the center of four first-layer molecules (phase I), or, at bridge-type positions between two first-layer TiOPc species (phase II). Our STM images unequivocally show that second-layer TiOPc molecules are located exclusively in the center of four first-layer TiOPc, i.e. indicating phase I type of stacking.

A quantitative analysis of the STM data (cf. Fig. 2c) yields an apparent height difference between the protrusions of first layer molecules and the aromatic plane of second layer molecules of about 1 Å. This is distinctly smaller than the height of the TiOPc molecule (~2.5 Å) and hence corroborates the interdigitated stacking of second-layer species.

**B    Vibrational signature of inhomogeneities**

Due to the distinct axial dipole moment of the titanyl group within the TiOPc molecule, in conjunction with the characteristic transformation of the axial Ti=O stretching mode $\nu_{Ti=O}$, it is convenient to use IRAS to monitor the evolution of this mode during the growth of the second TiOPc adlayer. In Fig. 3a a series of vibrational spectra describing the gradual modification of the Ti=O band with increasing coverages is displayed. At 1 ML, an intense and well defined Ti=O band is observed at 993.3 cm$^{-1}$, which reflects a uniform orientation of the Ti=O groups. Upon successively increasing the coverage of second-layer TiOPc, this band gradually vanishes and a sequence of four new Ti=O vibrational bands each shifted by about 10 cm$^{-1}$ to lower frequencies begin to grow at 981.4, 971.4, 961.4 and 956.9 cm$^{-1}$. It is apparent that the growth of the TiOPc bilayer proceeds via a number of well-defined intermediate structural configurations. Accompanied STM data (cf. Fig. 3c) show a non-continuous growth of the bilayer which instead consists of individual small islands formed by few bilayer molecules. According to the stepwise increase and decrease of frequency-shifted Ti=O bands we conclude that the mechanism causing this shift is acting locally and in a discrete rather than a continuous manner. Furthermore, its effect seems additive and the intensities of the frequency-shifted bands are directly related to the amount of second-layer TiOPc.

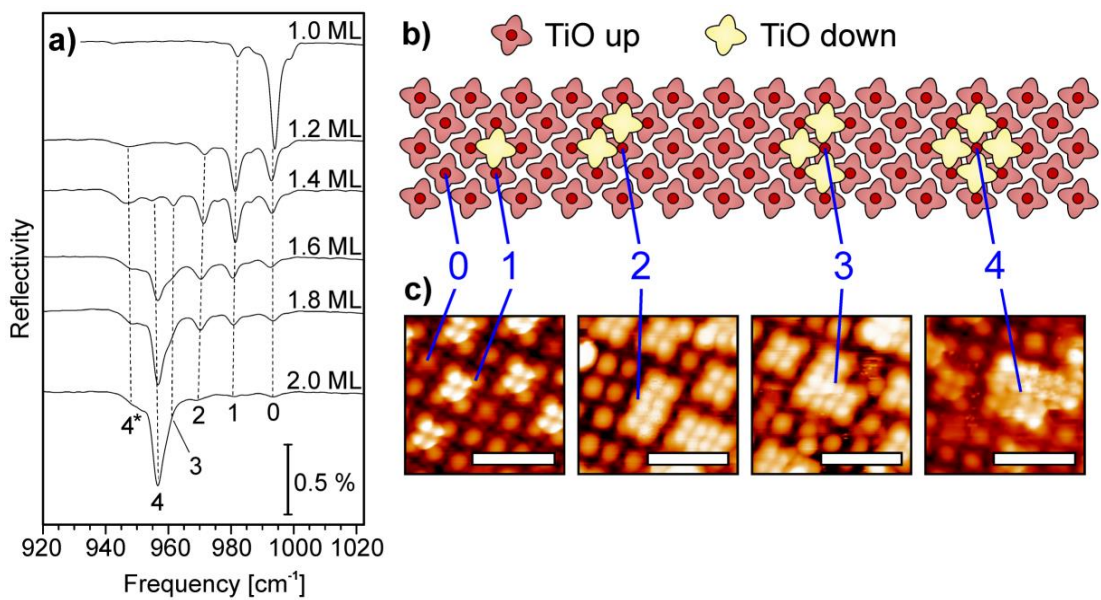

**Figure 3** Monolayer to bilayer transition of TiOPc on Ag(111). a) Coverage dependency of the out-of-plane Ti=O stretching mode; the IRAS spectra show a sequence of bands subject to gradual intensity variations and discrete shifts in frequency till completion of the second TiOPc layer. The spectra have been obtained at 80 K using a spectral resolution of 2 cm$^{-1}$. b) Schematic model describing local configurations of second-layer TiOPc molecules, encountered in the course of bilayer growth; the various local arrangements in particular explain the discreteness and the presence of four different frequency shifted bands. The numbers 0 till 4 identify the number of adjacent second-layer TiOPc neighbours (with O-down arrangement). c) STM micrographs ($U_{Bias}$ = 0.57 V, $I_t$ = 30 pA) depict examples of different coordinated monolayer molecules. Scale bars: 5nm.

A central question concerns the nature of the new bands and the origin of the discrete sequence of Ti=O bands upon growing TiOPc bilayers: Are they due to newly arriving second-layer TiOPc, or due to monolayer TiOPc which is subject to a specific intermolecular interaction that is causing this shift? As will be detailed in the following, the latter case is realized as it agrees in all respect with our experimental observations.

According to Fig. 3a, IRAS is very sensitive to changes in the local environment of the molecules and the presence of second-layer molecules is easily detected from the associated distinct and discrete frequency shifts of $v_{Ti=O}$. On the basis of our experimental observations we suggest a model describing the growth mechanism of the second TiOPc layer (Fig. 3b). It is apparent that each band between 993.3 and 956.9 cm$^{-1}$ is caused by a characteristic and discrete modification of the original Ti=O vibrational mode of the TiOPc monolayer species. Specifically, the line shifts are associated with a different number of adjacent neighbour molecules located in the second layer. The Ti=O vibrational band with the highest frequency (993.3 cm$^{-1}$) corresponds to the uncovered monolayer, i.e. with no second-layer molecules present (0 neighbours). The other modes correspond to the presence of one (981.4 cm$^{-1}$), two (971.3 cm$^{-1}$), three (961.4 cm$^{-1}$) or four (956.9 cm$^{-1}$) adjacent neighbours located in the second TiOPc layer. The detection of weak bands at 960 - 990 cm$^{-1}$ for the nominally saturated TiOPc bilayer points at residual areas with an incomplete second layer. These correspond to well-defined defects (i.e. vacancies of specific coordination) within the otherwise perfect bilayer which can occur e.g. at domain boundaries.

If the sequence of vibrational bands in Fig. 3a would originate from second-layer TiOPc, then the question arises what could have caused the sequential and distinct frequency shift of the Ti=O stretching mode, $v_{Ti=O}$, as the coverage increases. As the TiO sub-unit is located in the center of TiOPc, modifications at the periphery of the aromatic molecular frame (induced by e.g. the grouping of second-layer TiOPc, as observed in Fig. 3c) should affect this mode negligibly. Moreover, all second-layer TiOPc molecules experience an identical environment with respect to the underlying monolayer regardless of the bilayer coverage. It is thus much more likely that the sequentially shifted bands are due to monolayer TiOPc whose TiO group is interacting with one, two, three, or four second-layer molecules, depending on the actual coverage and local arrangement of second-layer TiOPc species, as depicted schematically in Fig. 3b.

For H-terminated (vicinal) Si(111) surfaces a similar sequence of discrete vibrational bands has been observed.[25] Specifically, each terrace consisting of a distinct number n of Si atom rows parallel to the straight step edges leads to discrete Si-H bands $v_n$, depending on the extent of dynamical dipole coupling. In fact, this correlation of individual terrace size and associated vibrational frequency shift may be interpreted as a quantum size effect for vibrational modes of interacting oscillators in discrete lattices.

Interestingly, the series in Fig. 3a shows a less intense but persistent vibrational band labelled 4* at 948.5 cm$^{-1}$. It is present right from the beginning of the series, i.e. at coverages slightly above 1ML, and it does not fit into the series of new peaks at 955 - 995 cm$^{-1}$. Its intensity initially grows proportional to the amount of second-layer TiOPc, but saturates (or even becomes weaker) upon completion of the bilayer. This band is ascribed to the Ti=O stretching mode of O-down oriented second-layer TiOPc. Its unusual intensity evolution behavior is associated with dynamical dipole coupling, in combination with the related intensity borrowing effect.[26-28] The intermolecular interaction thereby comprises coupling of the 4* band with similar modes of the TiOPc monolayer species. This coupling is more effective at close distances of the two interacting oscillators and if their vibrational frequencies differ only slightly. In our case the progressing redshift of the $\nu_{Ti=O}$ mode of monolayer species with increasing second-layer TiOPc coverage renders the intensity borrowing more and more effective. This causes a weakening of the 4* band at 948.5 cm$^{-1}$ in favour of the bands at higher frequencies which is in accordance with our observations. Confirmation for this assignment comes from the negligible frequency shift of this band as the bilayer growth proceeds. Second-layer TiOPc has its TiO group pointing downwards towards the center of four monolayer TiOPc species; its local environment does therefore not change in the course of the entire film growth sequence of Fig. 3a, in accordance with its negligible frequency shift.

There are two effects which, in our opinion, contribute to the particular stacking of TiOPc within the bilayer and also explain their thermal stability found in corresponding TDS experiments that are presented in the next section: (i) Due to the polar titanyl unit TiOPc molecules exhibit axial dipole moments which are oriented oppositely in the first and second layer and thus imply an attractive interaction for the laterally displaced dipoles. (ii) The small distance between peripheral hydrogen of second-layer TiOPc and the oxygen atom of the first-layer titanyl unit (and vice versa) promote the formation of hydrogen bonds. On the basis of the presently observed molecular stacking (cf. Fig. 2b) which parallels the structure of the TiOPc bulk phase I,[9] the latter was used to estimate the TiO - HC separation (indicated by green dashed line in Fig. 4). This yields a value of 2.3 Å. With regard to the enhanced thermal stability of bilayers compared to thicker films, a small interlayer contraction is anticipated so that this distance should rather be considered an upper limit. Following general trends,[29] formation of H-bonds of moderate strength is expected for such distances. Moreover, the interaction of the positively polarized H-C bond with the negatively polarized oxygen of the titanyl unit (cf. Fig. 4a), which is commonly considered as a proton acceptor, leads to a weakening of the Ti=O bond and thus a softening of the $\nu_{Ti=O}$ stretch mode which is in line with our experimental observation. Evidently this effect scales with the number of interacting bilayer molecules (i.e. the effective coordination of titanyl group, cp. Fig. 4b) and thus explains the step wise red-shift with increasing coverage.

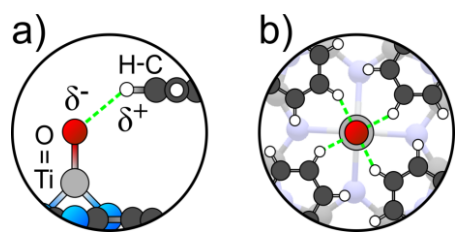

**Figure 4** Schematic illustration of the arrangement of first- and second layer TiOPc (sideview in (a) and top view in (b)); the suggested H-bonding between the oxygen of the TiO group and the hydrogens of the peripheral benzene units is indicated by the dashed lines.

## C. Thermal stability of the bilayer

The unusual thermal stability of the TiOPc bilayer as compared to thicker multilayers is demonstrated by our TDS data. As depicted in Figure 5 the bilayer desorbs at about 555 K while the bulk-like multilayer desorption signal occurs at about 495 K. This suggests a further interaction mechanism between the first and second layer. In addition to the above-mentioned stabilization effects by means of oppositely oriented dipoles and H-bonding between these species, an extra chemical interaction may occur between the downwards oriented oxygen of second-layer TiOPc and the Ag(111) surface. Evidently, this effect is restricted to the TiOPc bilayer only.

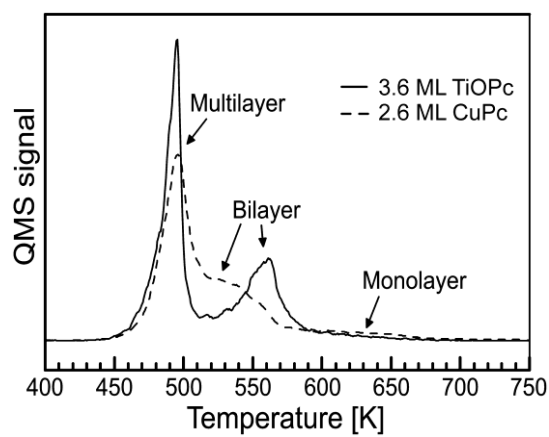

**Figure 5** Thermal desorption spectra (dT/dt = 1 K/s) of 3.6 ML of TiOPc (QMS mass 576 u) and 2.6 ML CuPc (QMS mass 575 u), both deposited onto a Ag(111) substrate at 300 K. The weak signal associated with monolayer desorption for both species is accompanied by incipient dissociation at T > 600 K.

The additional attractive forces not only influence the thermal stability of the bilayer but also their lateral packing. The corresponding SPA-LEED data (cf. Fig. 1) revealed a lateral compression yielding a reduction of the unit cell area from 199 Å$^2$ for the saturated monolayer to 189 Å$^2$ for the bilayer. Regarding thermal stability comparison with the related system CuPc/Ag(111), which exhibits neither an intrinsic dipole moment nor qualifies as a proton acceptor, yields a virtually identical desorption of multilayers, while CuPc bilayer films are stabilized only slightly (see Figure 5).

## D. From TiOPc bilayer to multilayer

Fig. 6 shows IRAS data of additional TiOPc layers grown on top of a TiOPc/Ag(111) bilayer. In the IR spectra distinct modes associated with the respective layers can be identified and assigned. From the displayed sequence of spectra it is apparent that the bilayer and thicker multilayer films can in fact be distinguished from their distinctly different spectral appearance. For example shifts in the line position of the prominent Ti=O stretching mode of about 10-15 cm$^{-1}$ are found for bi- and multilayers. This is attributed to the fact that molecules in the first two layers are still influenced by the underlying Ag(111) substrate: either via direct contact of the aromatic frame of monolayer TiOPc and the metal plane or due to interaction of the downward-looking oxygen of bilayer molecules with the metal substrate. While these additional stabilization does not occur in thicker films, such an interaction quite naturally explains the more pronounced frequency shift of the Ti=O stretching mode, located at ≈955 cm$^{-1}$ in bilayer films, as compared to ≈967 cm$^{-1}$ for thicker multilayers. Moreover, a significant screening of the TiO (oscillating) charge is anticipated, thereby rationalizing the weakness of the 4* band in Fig. 3a. The weak Ti=O sidebands at 978.3 and 986.4 cm$^{-1}$ are ascribed to non-ideally coordinated TiOPc within the (multiple) TiOPc bilayer stacks. At $\Theta_{TiOPc}$ > 4 ML no further frequency shifts are observed, most probably due to a well-defined bulk-like structure of the grown TiOPc films for all layers beyond the first bilayer, i.e. in the absence of additional substrate coupling. Comparison of 'in-plane' versus 'out-of-plane' mode intensities (vibrational bands in panel c and in panels a/b, respectively) confirm the more or less parallel stacking of TiOPc bilayers, even for thicker layers up to 12 ML (not shown). This conclusion notwithstanding, very weak bands, barely visible in 'low' coverage spectra ($\Theta_{TiOPc}$ ≈ 2-3 ML) become clearly discernible at higher $\Theta_{TiOPc}$ ≈ 6 ML. Conjectured variations in the tilt angle may be quantified by calculating the ratio of 'in-plane' versus 'out-of-plane' mode intensities. In a semi-quantitative evaluation we used the intensities of the modes at 1288 or 1393 cm$^{-1}$ and compared them to the prominent band at ≈737 cm$^{-1}$, or to $\nu_{Ti=O}$ to derive a measure for this ratio. According to our analysis the calculated ratio remains about constant in the range 2 - 12 ML (12 ML data not shown) which we take as strong evidence that the molecular plane does not start to tilt with increasing film thickness. We note that, similar to our arguments presented in section A, the observation of weak in-plane modes in our IR spectra cannot be taken as an indication or proof for a non-parallel arrangement of the TiOPc π-conjugated backbone since the slightly pyramidal molecular structure of TiOPc renders totally symmetric vibrational modes generally dipole active.

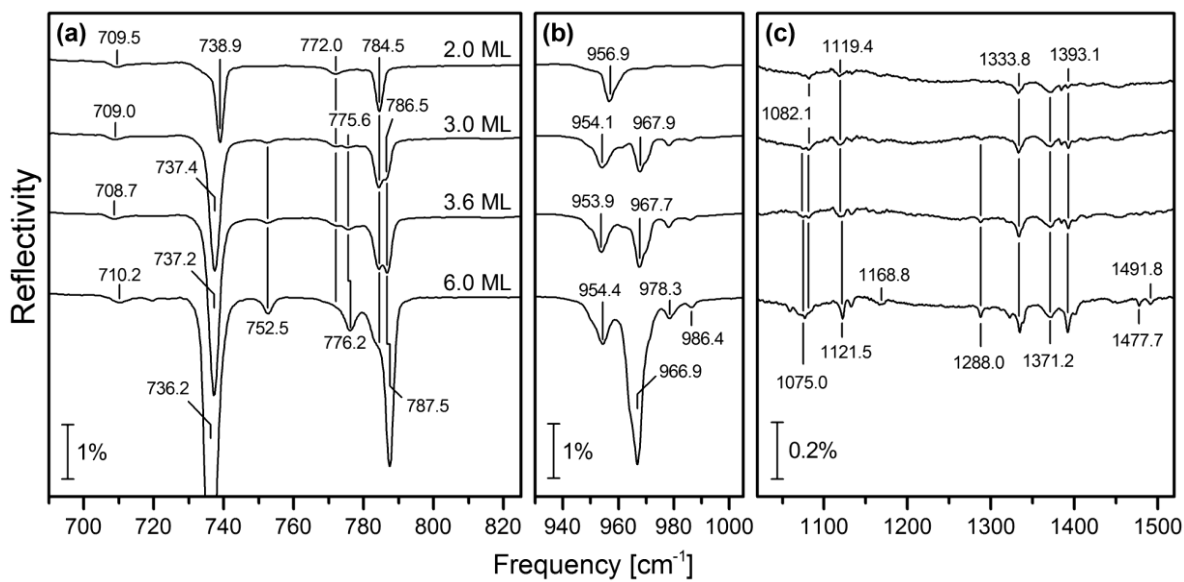

**Figure 6** Series of IR absorption spectra obtained from TiOPc layers with increasing nominal film thickness (coverage range 2 – 6 ML) adsorbed on Ag(111). TiOPc has been deposited at 300 K. The spectra in the panels (a-c) have been obtained from identical layers and the displayed spectral regions show the most prominent 'out-of-plane' C-H bending modes, (a), the Ti=O stretching mode, (b), as well as the various 'in-plane' deformation modes, (c). It is apparent that the 'in-plane' modes in general yield much lower intensities as compared to the C-H bending and Ti=O stretching modes in panels (a) and (b), rendering non-recumbent arrangements of the TiOPc within the bilayer planes unlikely. The spectra have been obtained at 80 K using a spectral resolution of 2 cm$^{-1}$.

## Conclusions

We have prepared and analysed the structure of TiOPc thin films on Ag(111) using different experimental techniques. Based on IRAS and STM data a microscopic model regarding the growth of the TiOPc bilayer structure is derived. According to our IR data and in accordance with STM results, the π-conjugated molecular frame of TiOPc, is oriented parallel to the Ag(111) surface for the mono-, bi- and multilayers films (coverage range 1 - 12 ML) and exhibits an alternating orientation of the titanyl units. We show that specific vibrations such as the Ti=O stretch provide detailed insight into the interface structure and binding situation. In particular, characteristic red-shifts are found which we attribute to H-bonding between peripheral hydrogen of bilayer TiOPc and oxygen of the monolayer TiOPc species. The actual amount of the red shift depends on the bilayer molecule coordination which in turn can be utilized to monitor the completion of bilayer films and the presence of vacancies. Since IRAS is also applicable to thicker films this facilitates the characterization of defect structures (vacancies) at buried interfaces. Moreover, we have demonstrated that both, the high intrinsic dipole moment of TiOPc, as well as the H-bond between oxygen and peripheral hydrogen of neighboring bilayer molecules, play a crucial role in the formation of the very first TiOPc layers. The unusual intrinsic stability of the TiOPc bilayer architecture suggests that this system can be combined with other π-conjugated molecules, e.g. to design molecular heterolayer structures with a stable and well defined interface.

## Acknowledgements

We gratefully acknowledge support from the Deutsche Forschungsgemeinschaft DFG (Germany) through the collaborative research center "Structure and Dynamics of Internal Interfaces" (SFB 1083, Projects A2, A3 and A7).